# Description of statistical switching in perpendicular STT-MRAM within an analytical and numerical micromagnetic framework

G. Siracusano[1,2], R. Tomasello[3], M. d'Aquino[4], V. Puliafito[5], A. Giordano[1], B. Azzerboni[5], P. Braganca[6], G. Finocchio[1,*], M. Carpentieri[7,*]

[1] *Department of Mathematical and Computer Sciences, Physical Sciences and Earth Sciences, University of Messina, I-98166 Messina, Italy*

[2] *Department of Computer Engineering and Telecommunications, University of Catania, I-95125, Catania, Italy*

[3] *Department of Engineering, Polo Scientifico Didattico di Terni, University of Perugia, I-50100 Terni, Italy*

[4] *Department of Engineering, University of Naples Parthenope, I-80143 Naples, Italy*

[5] *Department of Engineering, University of Messina, I-98166, Messina, Italy*

[6] *HGST, 3403 Yerba Buena Road, San Jose, CA 95035, United States*

[7] *Department of Electrical and Information Engineering, Politecnico di Bari, I-70125 Bari, Italy*

**Abstract**

The realistic modeling of STT-MRAM for the simulations of hybrid CMOS/Spintronics devices in comprehensive simulation environments require a full description of stochastic switching processes in state of the art STT-MRAM. Here, we derive an analytical formulation that takes into account the spin-torque asymmetry of the spin polarization function of magnetic tunnel junctions studying. We studied its validity range by comparing the analytical formulas with results achieved numerically within a full micromagnetic framework. We also find that a reasonable fit of the probability density function (PDF) of the switching time is given by a Pearson Type IV PDF. The main results of this work underlines the need of data-driven design of STT-MRAM that uses a full micromagnetic simulation framework for the statistical proprieties PDF of switching processes.

email: gfinocchio@unime.it, mario.carpentieri@poliba.it

## I. INTRODUCTION

The magnetization switching driven by spin-transfer torque (STT) [1,2,3,4] and spin-hall effect (SHE) [5,6,7,8,9,10] is the fundamental dynamics to design magnetic memory and spin-logic devices. Particularly, STT-driven magnetization is at the basis of emerging storage technology, i.e. STT-MRAMs [11,12], that are very attractive for their performances in terms of energy utilization, scalability and integration with complementary metal-oxide semiconductor (CMOS) process and technology [13]. The achievement of ultralow power consumption and storage scalability beyond CMOS is related to the advances in terms of materials and geometries. For instance, the use of materials with perpendicular magnetic anisotropy (PMA) permits to reach low critical current density (of the order of $10^6 A/cm^2$) and high thermal stability at the same time [14,15]. On the other hand, the realization of a reliable hybrid CMOS/Spintronics simulation environment is a key issue to address in order to obtain high performance of STT-MRAM [16,17]. Considering that STT-MRAM is going to be a mainstream of the storage industry, the implementation of a correct STT-MRAM model [18,19,20,21] becomes a central issue to simulate the behavior of hybrid devices combining CMOS technology and STT-MRAM. In particular, it is crucial for the design of writing and reading processes. The standard approach to model STT-MRAM is based on macrospin approximation [22,23,24,25], i.e. it is assumed that all the spins oscillate coherently during the switching process of the MTJ free layer and therefore a single domain is sufficient to describe the time evolution of the magnetization. However, since during STT-driven switching processes spatially non-uniform magnetization distribution can occur [26,27,28,29], the macrospin approach could become less accurate in describing the stochastic switching behavior of STT-MRAMs.

In this work, the stochastic switching in a perpendicular magnetic tunnel junction (MTJ) is studied by means of both a full micromagnetic model and an analytical formula derived from the macrospin approximation that, for the first time, takes into account the magnetization dependence of the STT efficiency. In order to achieve the analytical solution, in addition to the (*i*) macrospin approximation, the following hypotheses are necessary: (*ii*) in the absence of excitation, the energy barrier separating the equilibria is much higher than the thermal energy; (*iii*) the injected current is above the critical switching current of the device. We evaluate each of those hypotheses within the full micromagnetic model and highlight the main differences and limits in considering macrospin approach. We consider 1000 realizations that are sufficient to accurately model the basic mechanisms for switching to get the baseline switching probabilities for a given MRAM stack/geometry. These probabilities to first order allow to understand distributions and that is very significant and could be useful in future improvements.

A further important result is that, while Zhao *et al.* [19] found that an asymmetric probability density function (PDF) of the switching time is well-reproduced by skew normal distribution, here we show that, in order to achieve a cumulative quadratic error at least one order of magnitude lower than the skew normal distribution, the PDF should follow a Pearson Type IV function.

The paper is organized as follow. Section II discusses the details of the micromagnetic framework and the mathematical formulation of the Pearson Type IV PDF. The description of the results and discussion are presented in Sections III and IV while conclusions are summarized in Section IV.

## II. MICROMAGNETIC FRAMEWORK

*(a) Device and full micromagnetic model*

We study a state of the art MTJ with a CoFeB(1 nm) Free Layer (FL) and a circular cross section (diameter of 30 nm) within a full micromagnetic framework [30,31,32]. The Landau-Lifshitz-Gilbert-Slonczewski (LLGS) equation in dimensionless form can be written as [33]:

$$\frac{\partial \mathbf{m}}{\partial \tau} = -\mathbf{m} \times \left[ \mathbf{h}_{\text{eff}} - \alpha \frac{\partial \mathbf{m}}{\partial \tau} - \beta \frac{\mathbf{m} \times \mathbf{m}_\mathbf{P}}{1 + c_p \mathbf{m} \cdot \mathbf{m}_\mathbf{P}} + \mathbf{h}_{\text{th}} \right] \quad (1)$$

where $\alpha$ is the Gilbert damping, $\mathbf{m}$ and $\mathbf{m_P}$ are the normalized magnetization of the FL and pinned layer respectively. $\tau = \gamma_0 M_s t$ is the dimensionless time, with $\gamma_0 = 2.21 \times 10^5$ m(As)$^{-1}$ being the absolute value of the gyromagnetic ratio, and $M_s$ the saturation magnetization. $\mathbf{h}_{\text{eff}}$ is the normalized effective field that includes the exchange, magnetostatic, anisotropy and external fields, and the Oersted field due to the current. $\mathbf{h}_{\text{th}}$ is the thermal field given by $\mathbf{h}_{\text{th}} = \chi \sqrt{2\alpha k_B T / (\mu_0 M_S^2 V)} = \nu \chi$, where $k_B$ is the Boltzmann constant, $\mu_0$ is the vacuum permeability, $V$ is the volume of the free layer, $T$ is the temperature of the sample and $\chi$ is a three-dimensional isotropic white Gaussian noise uncorrelated in space and time [34,35,36]. The intensity $\nu = \sqrt{(2\alpha K_B T)/(\mu_0 M_s^2 V)}$ of the thermal fluctuations is obtained from the fluctuation-dissipation theorem [33,37,38] and does not depend on the numerical discretization of Eq. (1) [36]. $\beta$ is the normalized injected current density given by $\frac{2\eta g |\mu_B| J_{MTJ}}{e \gamma_0 M_S^2 t_{FL}}$ [39,40] where $g$ is the Landè factor, $\mu_B$ is the Bohr magneton, $e$ is the electron charge, $t_{FL}$ is the thickness of the FL and $\eta$ is the spin-polarization factor [39], $c_p = \eta^2$ defines the spin-torque asymmetry. $J_{MTJ}$ is the current density flowing perpendicularly through the whole FL cross section. The parameters used in this study are: $M_S$=1000 kA/m [14], exchange constant $A$=20 pJ/m [41], perpendicular anisotropy constant $k_u$=0.80 MJ/m$^3$, and $\alpha$=0.03 [14]. We consider a synthetic antiferromagnetic polarizer with a reference magnetization pointing along the positive out-of-plane direction $\mathbf{m}_p = \mathbf{m}_z$ and a negligible dipolar coupling. No external field is considered in our study. For the micromagnetic numerical solution of Eq. (1), the magnetic system is discretized into a mesh of cubic cells of 1nm × 1nm × 1nm.

*(b) Analytical solution for the probability density function*

In order to derive an analytical formulation for the probability density function (PDF) of switching times, we assume three main hypotheses. (*i*) The magnetization is spatially-uniform during the dynamics, so that the magnetic particle can be described within a macrospin approximation. Under this assumption, the effective field is given by:

$$\mathbf{h}_{eff} = -\frac{\partial g_L}{\partial \mathbf{m}} \quad , \quad g_L(\mathbf{m}, \mathbf{h}_a) = \frac{1}{2} D_x m_x^2 + \frac{1}{2} D_y m_y^2 + \frac{1}{2} D_z m_z^2 \tag{2}$$

where $D_x$, $D_y$, $D_z$, are effective anisotropy factors [42].

(*ii*) In the absence of excitation, the energy barrier separating the equilibria is much higher than the thermal energy. This hypothesis is suitable at large enough perpendicular anisotropy and it is valid for the parameters used in this work. (*iii*) The injected current is above the critical switching current of the device. In this situation, the switching time can be evaluated considering the deterministic magnetization motion acting on a random initial magnetization distribution due to thermal fluctuations [43,44]. In the absence of current, the magnetization is distributed according to the stationary solution of the Fokker-Planck equation [38,42], which, in terms of small tilting angle $\theta$ ($\sin\theta \approx \theta$) with respect to the symmetry out-of-plane axis, can be expressed in the limit of the hypothesis (*ii*) as:

$$p_{eq} = \mu k_{eff} \theta \exp\left[-\mu \frac{k_{eff}}{2} \theta^2\right] \tag{3}$$

where $k_{eff} = D_\perp - D_z = 0.412$ and $\mu = \frac{2\alpha}{\nu^2}$. Here, we consider $D_\perp = D_x = D_y = 0.0464$ (as computed micromagnetically) and $D_z = -0.366$ being the device with a circular cross section and asymmetric spin-torque ($c_p = 0.66^2 = 0.436$) [45,46]. By neglecting thermal fluctuations in Eq. (1), namely setting $\nu = 0$, it is possible to show that the threshold switching current is (detailed derivation will be presented elsewhere):

$$\begin{aligned}\beta_{crit}^{P \to AP} &= \alpha(1+c_p)(k_{eff}) \\ \beta_{crit}^{AP \to P} &= -\alpha(1-c_p)(k_{eff})\end{aligned} \tag{4}$$

For the magnetization switching from parallel to anti-parallel state (P→AP) and vice versa (AP→P), the switching times $t_{s,P \to AP}$ / $t_{s,AP \to P}$ are considered as the time interval between the application of the current (the initial z-component of the magnetization is $m_{z0}$) and the time instant where the z-component of the magnetization is equal to $m_{zf}$=-0.9 / 0.9 (<$m_z$>=-0.9 / 0.9 in the case of full micromagnetic simulations). For P→AP switching, when $\beta > \beta_{crit}^{P \to AP}$ the deterministic $t_{s,P \to AP}$ is given by:

$$t_{s,P \to AP} = \frac{1}{\alpha} G(m_{z0}, m_{zf} = -0.9) = g(m_{z0}) \quad . \tag{5}$$

where the function $G(m_{z0}, m_{zf})$ is the (dimensionless) time required for magnetization to reach $m_z = m_{zf}$ starting from the initial state $m_z = m_{z0}$. The function $g(m_{z0})$ in Eq. (5) can be represented in closed form by integrating Eq. (1) within the hypotheses (*i*), (*ii*) and (*iii*) by using the separation of variables and standard integrals (the calculations are a bit lengthy but straightforward). The result is:

$$t_{s,P \to AP} = -\frac{1}{2\alpha} \frac{\begin{bmatrix} k_{eff}C_4C_1 - k_{eff}C_5C_1 - 2C_3k_{eff}^2 - 6C_3\xi c_p k_{eff} \\ +2C_3k_{eff}^2 c_p^2 + 6C_2\xi c_p k_{eff} + 2C_2k_{eff}^2 - 2C_2k_{eff}^2 c_p^2 \\ +k_{eff}c_p^2 C_5C_1 - c_p C_5\xi C_1 - k_{eff}c_p^2 C_4C_1 + c_p C_4\xi C_1 + \\ +\ln\left[\frac{(1-m_{zf}^2)}{(1-m_{z0}^2)}\right]^{c_p^2 C_1 k_{eff} - c_p C_1\xi - C_1 k_{eff}} + \ln\left[\frac{(1+m_{zf})(1-m_{z0})}{(1+m_{z0})(1-m_{zf})}\right]^{C_1\xi} \end{bmatrix}}{\left[(k_{eff}c_p - \xi + k_{eff})(-k_{eff} - \xi + k_{eff}c_p)C_1\right]} = g(m_{z0})$$

(6)

where the following quantities have been defined for sake of compactness:

$$\begin{aligned} \xi &= \beta/\alpha \\ C_1 &= \sqrt{4\xi c_p k_{eff} - 2h_{az} c_p k_{eff} + k_{eff}^2} \\ C_2 &= \operatorname{arctan} h\left(\frac{k_{eff} + 2k_{eff}c_p m_{z0}}{C_1}\right) \\ C_3 &= \operatorname{arctan} h\left(\frac{k_{eff} + 2k_{eff}c_p m_{zf}}{C_1}\right) \\ C_4 &= \ln\left|k_{eff}m_{zf} + k_{eff}m_{zf}^2 c_p - \xi\right| \\ C_5 &= \ln\left|k_{eff}m_{z0} + k_{eff}m_{z0}^2 c_p - \xi\right| \end{aligned} \tag{7}$$

By combining Eqs. (5)-(7) with Eq. (3), the PDF $f_{t_s,P \to AP}$ and cumulative distribution function (CDF) $F_{t_s,P \to AP}$ for the stochastic switching time $t_{s,P \to AP}$ can be calculated by the following expressions. In this respect, the former is given by:

$$f_{t_s,P\to AP}(t_{s,P\to AP}) = \mu k_{eff} \exp\left[-\mu \frac{k_{eff}}{2} \arccos^2\left(g^{-1}(t_{s,P\to AP})\right)\right] \cdot$$
$$\left|\alpha\left[k_{eff} g^{-1}(t_{s,P\to AP}) - \beta/\alpha\left(1+c_p g^{-1}(t_{s,P\to AP})\right)^{-1}\right]\left[1-\left(g^{-1}(t_{s,P\to AP})\right)^2\right]\right|,$$

(8)

and the latter is:

$$F_{t_s,P\to AP}(t_{s,P\to AP}) = \exp\left[-\mu \frac{k_{eff}}{2} \arccos^2\left(g^{-1}(t_{s,P\to AP})\right)\right]$$

(9)

In Eqs. (8)-(9), the function $g^{-1}(t_s)$ cannot be written in closed form, but nevertheless can be efficiently obtained by numerical inversion of the strictly monotone function $g(m_{z0})$ expressed by Eqs. (6)-(7). In the above expression the switching mechanism from parallel to anti-parallel state, P→AP, is considered. Similar equations can be derived for AP→P switching. We remark that the above formulation extends the analytical theory [43, 44] to devices with asymmetric spin-torque such as MTJs.

*(c) Statistical description of the micromagnetic simulations at T=300K*

The switching process of the magnetization is strongly nonlinear, thus the PDF of the $t_s$ is expected to be non-Gaussian. While in Ref. [19], the skew normal distribution was used to describe the asymmetry of the PDF of $t_s$, here we show that the kurtosis is necessary to fit the micromagnetic PDFs of $t_s$. Among the Pearson distribution function types (I–XII) [47,48,49], we find that the Pearson Type IV (p4PDF) [50,51,52] (an asymmetric version of the Student's *t* distribution [53]) is the best option for our framework. By considering the first four standardized moments mean $\mu$, standard deviation (STD) $\sigma$, skewness $\gamma$ and kurtosis $\beta_2$, the p4PDF can be written as:

$$f(x|\mu,\sigma,\gamma,\beta_2) = \frac{k}{\sigma}\left[1+\left(\frac{\hat{x}-\lambda}{a}\right)^2\right]^{-m} \exp\left[-\upsilon \tan^{-1}\left(\frac{\hat{x}-\lambda}{a}\right)\right] \quad (m>1/2)$$

(10)

$$\lambda = \frac{av}{b}$$

$$v = \frac{2c_1(1-m)}{\sqrt{4c_0 c_2 - c_1^2}}$$

$$\beta_1 = \gamma^2$$

$$c_0 = \frac{4\beta_2 - 3\beta_1}{D}$$

where $\hat{x} = \frac{x-\mu}{\sigma}$, whereas $a = \sqrt{\frac{b^2 + (b-1)}{b^2 + v^2}}$, and $c_1 = \frac{\gamma \cdot (\beta_2 + 3)}{D}$ (11)

$$b = 2m - 1$$
$$m = 1/2c_2$$

$$c_2 = \frac{\gamma \cdot (2\beta_2 - 3\beta_1 - 6)}{D}$$

$$D = (10\beta_2 - 12\beta_1 - 18)$$

In Eq. (11) *m*, *v*, *a*, and *λ* are real-valued parameters, and $-\infty < x < \infty$. *k* is a normalization constant that depends on *m*, *v*, *a*, and can be expressed by:

$$k = \frac{\left| \frac{\Gamma(m + (v/2) \cdot 1i)}{\Gamma(m)} \right|^2}{a \cdot \beta\left(m - \frac{1}{2}, \frac{1}{2}\right)} \quad (12)$$

Here, $\Gamma(m)$ is the gamma function defined as $\Gamma(m) = \int_0^\infty e^{-t} t^{m-1} dt$ and $\beta(z,w)$ is the beta function defined as $\beta(z,w) = \int_0^1 t^{(z-1)} \cdot (1-t)^{(w-1)} dt = \frac{\Gamma(z)\Gamma(w)}{\Gamma(z+w)}$ [54]. Such computations have been performed within a parallel processing framework, which has been designed and implemented for accelerating algorithms computation [55, 56].

### III. STATISTICAL SWITCHING IN PERPENDICULAR STT-MRAM

In the rest of the work, we will focus on the P→AP switching process (positive current densities applied), while qualitative similar results are valid for AP→P switching. Fig. 1(a) shows a comparison between the $t_{s,P \to AP}$ as a function of the current density (critical current density $J_{Crit}$=2.7 MA/cm$^2$ to achieve a fast switching process, where the switching time is shorter than about 10 ns) as computed from micromagnetic simulations (solid blue curve with circles) and Eq. (6) (solid red curve) for *T*=0 K. The good agreement is achieved because, without thermal fluctuations, the switching process in the full micromagnetic framework can be well described by the macrospin approximation (see Fig. 1(b) for an example of the switching trajectory of the normalized magnetization <*m*> at $J_{MTJ}$=5.0 MA/cm$^2$).

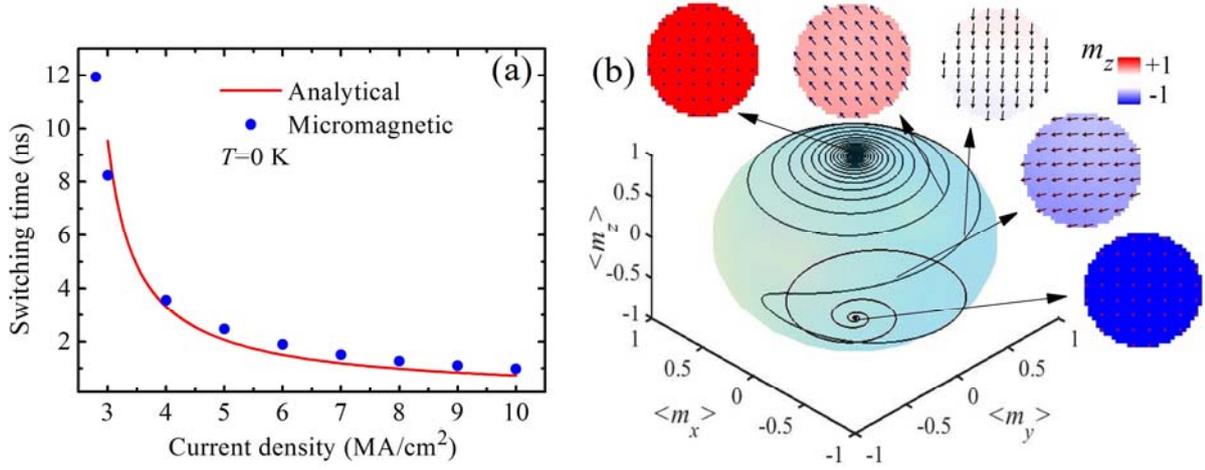

FIG. 1: (a) Switching time as a function of the current density obtained from the analytical (red line) and micromagnetic computations (blue line with circles) at $T$=0 K. (b) Switching trajectory obtained from micromagnetic simulations at $J_{MTJ}$=5.0 MA/cm$^2$ with some representative magnetization snapshots displayed as insets (red positive, blue negative out-of-plane component of the magnetization).

Figs. 2 (a-c) and (d-f) report the PDFs and the cumulative distribution functions (CDFs) of the switching time $t_{s,P \to AP}$ for three current densities (a) $J_{MTJ}$=3 MA/cm$^2$, (b) $J_{MTJ}$=6 MA/cm$^2$, and (c) $J_{MTJ}$=10 MA/cm$^2$ as predicted from the analytical theory (red lines), computed from micromagnetic simulations (blue symbols in Figs. 2(a-c), blue lines in Figs. 2(d-f)) and from the fitting with the p4PDF (solid black lines) and the skew normal (solid green lines). Similar simulations at $T$=0 K and $T$=300 K have been performed by including the perpendicular torque contribution $q(V)(\mathbf{m} \times \mathbf{m_P})$ into Eq. (1), with $q(V)$ being the voltage dependent parameter [40,57,58,59] (see Appendix A). No significant changes were observed on mean switching time and STD, as well as on the CDFs, leading to the conclusion that the perpendicular torque contribution is negligible.

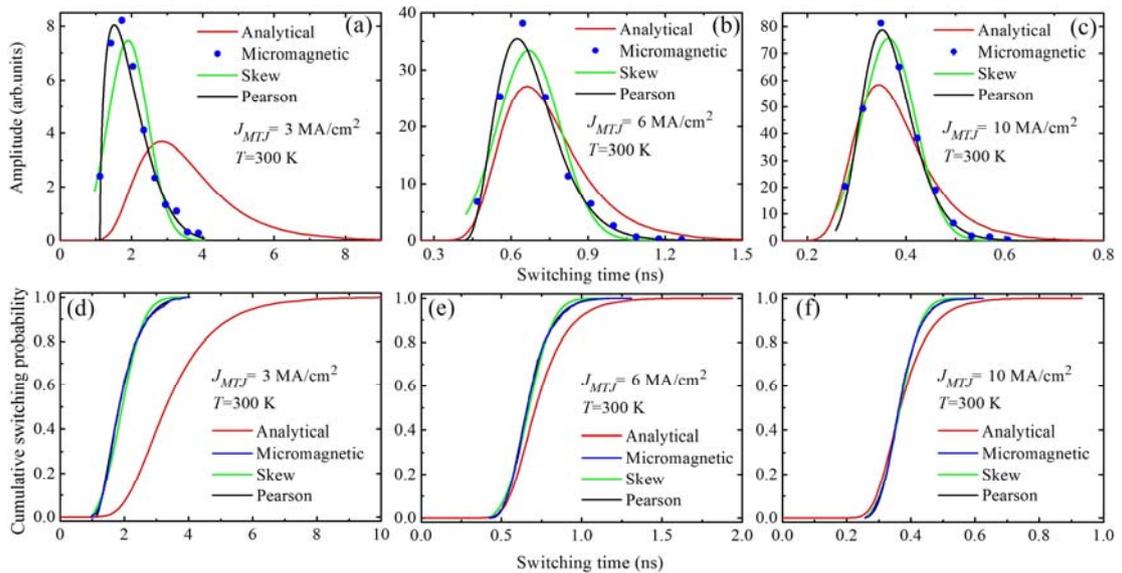

FIG. 2: Probability distribution function computed by means of analytical model, micromagnetic simulations (blue circles obtained from the histogram of simulations), Pearson (p4PDF) and skew normal approximation using simulations data at room temperature for (a) $J_{MTJ}$=3 MA/cm$^2$, (b) $J_{MTJ}$=6 MA/cm$^2$, (c) $J_{MTJ}$=10 MA/cm$^2$. Cumulative distribution function computed by means of analytical model (solid red line), empirical CDF using micromagnetic simulations (solid blue line), Skew normal CDF (solid green line) and Pearson CDF (solid black line) using simulations data at room temperature for (d) $J_{MTJ}$=3 MA/cm$^2$, (e) $J_{MTJ}$=6 MA/cm$^2$, (f) $J_{MTJ}$=10 MA/cm$^2$.

Previous measurements [19] have shown that PDFs of switching process (10.000 realizations) in in-plane MTJ with elliptical cross section (50x150nm$^2$) are well approximated by a skew normal distribution. However, here we find that also the kurtosis (p4PDF) has to be taken into account to have a good approximation of the micromagnetic PDF (we focused on fast switching processes where the $t_{s,P \to AP}$ is shorter than 10ns). In other words, for the whole current region, the p4PDF reproduces the micromagnetic data with a significant improvement in terms of error over the skew normal distribution. In this concern, we perform a quantitative analysis of the cumulative quadratic error $CQ_{err}$ computed from the following expressions:

$$CQ_{err} = \sum_{i=1}^{N-1} \Delta t_{(i,j)} \cdot \left( \Delta \hat{F}_e(i,j) - \Delta \Psi(i,j) \right)^2, (j = i+1) \tag{13}$$

where $\Delta \hat{F}_e(i,j) = \hat{F}_e(t_j) - \hat{F}_e(t_i)$, and

$$\hat{F}_e(t_s) = \frac{1}{N} \sum_{i=1}^{N} 1_{xi \leq t_s} = \frac{1}{N} \sum_{i=1}^{N} number\ of\ elements\ x_i\ :\ t_{|xi} \leq t_s$$

with $\Delta \Psi(i,j)$ defined as

$$\Delta \Psi(i,j) = \Psi(j) - \Psi(i), where\ \Psi \begin{cases} Normal\ CDF \\ Skew\ normal\ CDF \\ Pearson\ CDF \end{cases} \tag{14}$$

where $\hat{F}_e$ is the empirical CDF of the available simulation data ($N$ realizations), $t_s$ is the switching time ($t_{s,P \to AP}$ for Fig. 3) as extracted from simulations, and $\Psi$ represents a given CDF among the ones as shown in the same expression.

Fig. 3 reports the $CQ_{err}$ as a function of the current density $J_{MTJ}$ as extracted from Normal (red line with dots), Skew Normal (green line with circles) and Pearson (black line with triangles) CDFs, respectively. The largest error is given by the Normal CDF because the data do not exhibit a Gaussian-like shape. The skew normal CDF [19] reduces the $CQ_{err}$ by one order of magnitude whereas the Pearson CDF (that accounts for both skewness and kurtosis) provides a $CQ_{err}$ at least two orders of magnitude smaller than the one of the Normal distribution over the whole range of current values.

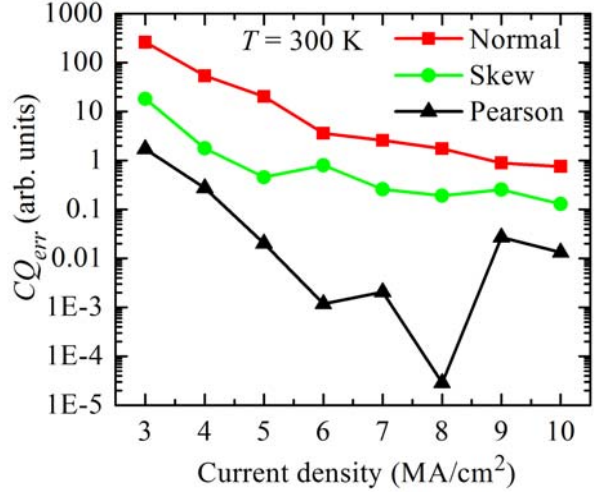

FIG. 3: Cumulative quadratic error $CQ_{err}$ as a function of the current density $J_{MTJ}$ as extracted from Normal CDF (red line with squares), Skew Normal CDF (green line with circles) and Pearson CDF (black line with triangles).

This result has the following implication for the modeling of the switching process in STT-MRAM. When the PDF associated to a given scenario has a non-trivial statistical behavior, the traditional approaches based on Gaussian approximation [60] or recent Skew normal distribution generalization [19] cannot be sufficiently accurate and then to this end, the application of Pearson distribution family is necessary to attain a better accuracy. This achievement is important to take into account for hybrid CMOS-STT-MRAM simulation environments.

Figs. 2 (a-c) and (d-f) also show the analytical PDFs and CDFs, respectively. In particular, it can be observed that the analytical results are close to the micromagnetic ones at moderate and large current densities, consistently with the assumptions made in section II(b).

Fig. 4 summarizes the four statistical moments of the switching time ((a) mean, (b) STD, (c) skewness and (d) kurtosis) as a function of the current density.

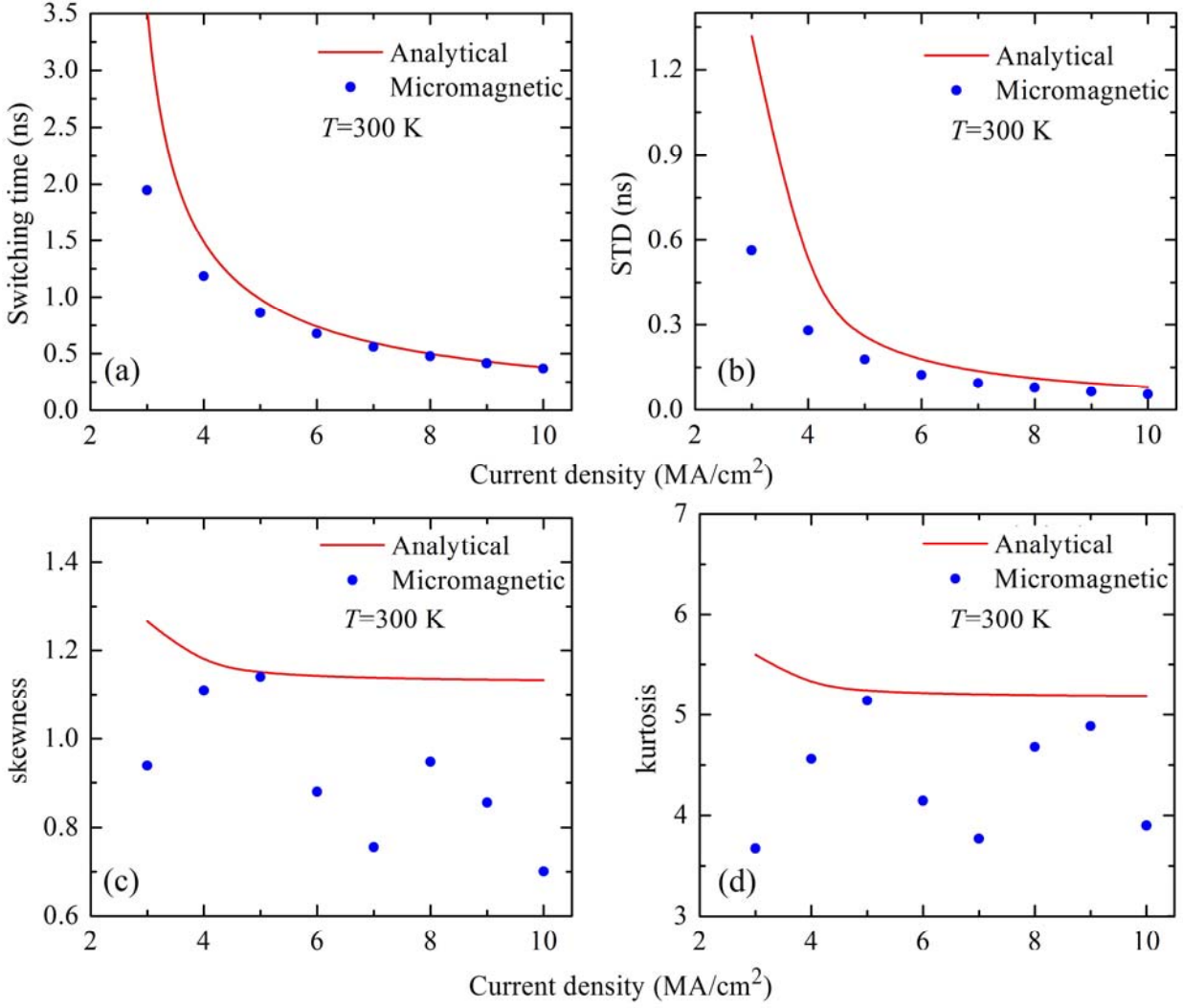

FIG. 4: (a) Switching time as a function of the current density obtained from the analytical (red line) and micromagnetic computations (blue line with circles) at $T$=300 K. (b) Standard deviation, (c) skewness and (d) kurtosis as a function of the current density $J_{MTJ}$ as computed by using analytical (red squares) and micromagnetic model (blue circles), respectively.

A comparison between the $t_{s,P \to AP}$ and the STD (Fig. 4(a-b)) shows a disagreement at current below 5.0 MA/cm$^2$. On the other hand, skewness and kurtosis (Fig. 4 (c-d)), exhibit both a quantitative and qualitative difference in the whole range of current.

In order to understand the differences between analytical and micromagnetic results, we have checked all the hypothesis of the analytical model within the full micromagnetic scenario. We have started from the approximation $\sin\theta \approx \theta$ used in Eq. (3). Fig. 5(a) represents the cone of the magnetization distribution for the $P$ state, the maximum angle of 15.48° gives rise to an error smaller than 5%. Therefore, this hypothesis is valid also within the micromagnetic framework.

The analytical model neglects the stochastic thermal fluctuations during the switching process (hypothesis (*iii*)), entailing that, once the random initial state (IS) of the magnetization is known, then the switching time can be deterministically calculated. With this in mind, within our micromagnetic framework, we have performed simulations considering a fixed IS as well as the stochastic thermal contribution during the switching mechanism. Fig. 5(b) shows the p4PDF for three current densities ($J_{MTJ}$=3, 6, and 10 MA/cm$^2$). As the current increases, the p4PDF becomes sharper, i.e. the STD decreases, leading to a more deterministic switching process. Our results showing that this hypothesis is valid for currents well above the critical switching current (>1.5$J_{Crit}$), are also in agreement with previous experiments [61] and numerical studies [62,63,64,65]. We conclude that this is the reason why the first and second order analytical moments fit well at high currents.

We have also checked the hypothesis of macrospin approximation during the switching. Fig. 5(c) illustrates the uniformity degree (UD) as computed during the switching process at zero and room temperature [66]:

$$UD = 1 - \sqrt{\frac{\left(\langle m_x^2 \rangle - \langle m_x \rangle^2\right) + \left(\langle m_y^2 \rangle - \langle m_y \rangle^2\right) + \left(\langle m_z^2 \rangle - \langle m_z \rangle^2\right)}{3}} \qquad (15)$$

where $\langle m_i \rangle$ is the normalized magnetization (*i*=*x,y,z*) averaged over all the free layer cells. As mentioned above, at zero temperature, the switching dynamics can be approximated within a macrospin model ($UD \approx 1$), whereas, the stochastic thermal fluctuations at room temperature lead to a less uniform switching mechanism ($UD$ drops to less than 0.80). We conclude that this aspect influences the high order statistical moments of the PDF and gives rise to the difference in the skewness and kurtosis between analytical and micromagnetic results in the whole range of current. The macrospin approximation is no longer valid if the interfacial Dzyaloshinskii–Moriya interaction is taken into account (see Appendix B).

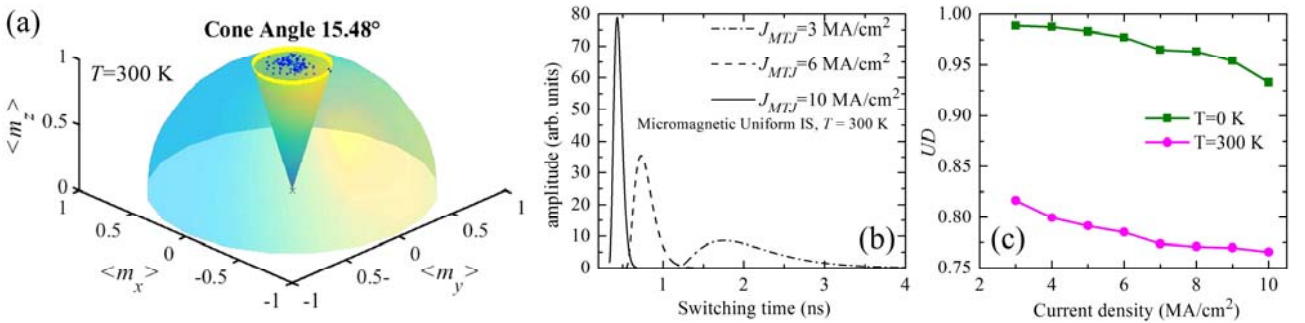

FIG. 5: (a) Deviation cone angle calculated from the initial random distribution of the magnetization at room temperature and $J_{MTJ}$=0 MA/cm$^2$. (b) p4PDF calculated at room temperature for $J_{MTJ}$=3 MA/cm$^2$ (dash-dot line), $J_{MTJ}$=6 MA/cm$^2$ (dot line), and $J_{MTJ}$=10 MA/cm$^2$ (solid line), when the initial configuration is the uniform magnetization state and the stochastic thermal fluctuations are considered only during the switching process. (c) Uniformity degree (UD) as a function of the current density at zero (green line with squares) and room temperature $T$=300 K (magenta line with circles).

## IV. CONCLUSIONS

In this paper, the STT switching distribution in a circular MTJ cell has been studied by characterizing the statistical properties of switching PDF and CDF by means of extensive numerical simulations (1000 iterations). We have found that Pearson Type IV PDF fits well the statistics of the switching process as computed from micromagnetic simulations. In addition, an analytical model that correctly describes the switching distribution for current regimes higher than 1.5$J_{Crit}$ has been developed. We tested the validity range of the analytical model by numerical simulations and we concluded that a micromagnetic model approach is necessary to predict skewness and kurtosis at both low and high current regimes. On the other hand, for current values larger 1.5 times than the threshold one, the macrospin approach is sufficient to describe the mean and standard deviation of the numerical PDF. In conclusion, our work underlines the need of data-driven design of STT-MRAM cells based on combined analytical and full micromagnetic approaches, highlighting the physical aspects related to these kind of phenomena, and providing physical results useful for engineers.

## ACKNOWLEDGEMENTS

This work was supported by the project PRIN2010ECA8P3 from Italian MIUR. R. T. acknowledges Fondazione Carit - Projects – "Sistemi Phased-Array Ultrasonori", and "Sensori Spintronici". M.d'A. acknowledges Programme for the Support of Individual Research 2015-2017 - year 2015 - funded by University of Naples "Parthenope". The simulations have been performed with the availability of SCL (Scientific Computing Laboratory) of the University of Messina, Italy.

# APPENDIX A: Effect of the field like torque

We performed numerical simulations by including the field like torque contribution. Eq. (1) of the main text now reads [67]:

$$\frac{\partial \mathbf{m}}{\partial \tau} = -\mathbf{m} \times \left\{ \mathbf{h}_{eff} - \alpha \frac{\partial \mathbf{m}}{\partial \tau} - \frac{\beta}{1 + c_p \mathbf{m} \cdot \mathbf{m}_P} \left[ \mathbf{m} \times \mathbf{m}_P - q(V)\mathbf{m}_P \right] + \mathbf{h}_{th} \right\} \quad (A1)$$

where we consider $q(V) = aV^2$, with $a$ being the parameter linking the field like torque to the voltage square $V^2$, and can be identified from experimental measurements [40,57,58]. We are taking into account only the term proportional to the square voltage because we are focusing on a symmetric system [59], while the term linear with $V$ is significant in the case of an asymmetric MTJ [57]. As experimental determined in Ref. [40], we have considered a variation of the field like torque from 0 to 25% of the damping torque. Fig. A1 shows the switching times ($T$=0 K) as a function of the current density $J_{MTJ}$ for different values of the field like torque (P->AP switching). As can be observed, a negligible variation of the switching time on $q(V)$ is obtained.

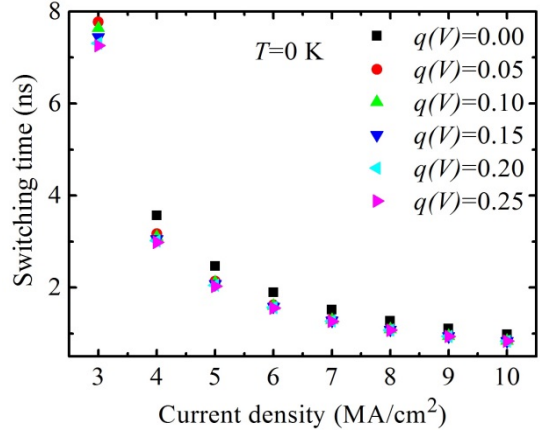

FIG. A1: Switching time as a function of the current density for different values of the voltage dependent parameter $q(V)$.

The second part of our test is based on running simulations at room temperature ($T$=300 K, 1000 iterations) for two different current density values ($J_{MTJ}$=3.0 MA/cm² that is near the threshold, and $J_{MTJ}$=10.0 MA/cm²), and for three values of $q(V)$ equal to 0.10 and 0.25 (the value $q(V) = 0.0$ has been already considered in the main text of the manuscript). The results are summarized in Figs. A2(a) and (b) that show the cumulative switching probability for (a) $J_{MTJ}$=3.0 MA/cm² and (b) 10.0 MA/cm², respectively. In A2(c) and A2(d), we show the percentage difference for the four statistical moments computed as $percentage = \frac{s_q(i) - s_0(i)}{s_0(i)} \bigg|_{i=mean,std,skewness,kurtosis}$, where $s_q(i)$ and $s_0(i)$ are the statistical moments calculated for a finite $q(V)$ and $q(V)=0.0$, respectively. Our results indicate that there is no qualitative change in presence of field-like torque and that the Pearson PDF function should be still used to achieve a cumulative quadratic error $CQ_{err} \approx 10^{-3}$ (see Eqs. (13) and (14) of the main text). In other words, the Pearson PDF gives the best fitting even in presence of the field like torque contribution, thus no qualitative differences are introduced in the PDFs in presence of the field-like torque.

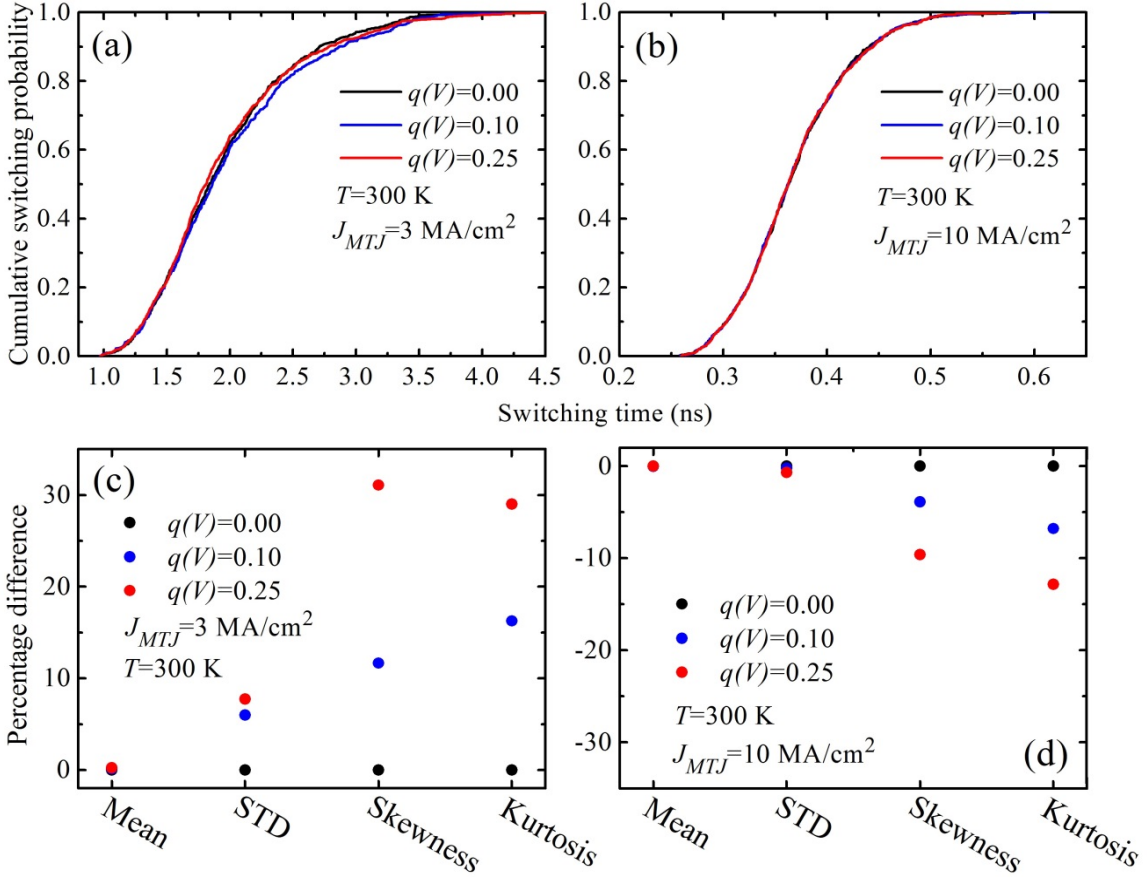

FIG. A2: Cumulative switching probability for different values of the perpendicular torque for (a) $J_{MTJ}$=3.0 MA/cm² and (b) $J_{MTJ}$=10.0 MA/cm². Percentage difference of the mean switching time, STD, skewness, and kurtosis with and without $q(V)$ for current density (c) $J_{MTJ}$=3.0 MA/cm² and (d) $J_{MTJ}$=10.0 MA/cm².